\documentstyle[pre,aps,twocolumn]{revtex}
\begin{document}
\draft
\tighten

\title{Isotropic phase-number squeezing and macroscopic
           quantum coherence}

\author{G. M. D'Ariano\cite{dar}}
\address{Dipartimento di Fisica ``Alessandro Volta'', Universit\`a di Pavia,
         Via A. Bassi 6, I--27100 Pavia, Italy}
\author{M. Fortunato\cite{for}}
\address{Dipartimento di Fisica, Universit\`a di Roma ``La Sapienza'',
         P.le A. Moro 2, I--00185 Roma, Italy \\ {\rm \ } \\ {\rm and} }
\vskip -.3cm
\author{P. Tombesi\cite{tom}}
\address{ Dipartimento di Matematica e Fisica, Universit\`a di Camerino,
          Via Madonna delle Carceri, I--62032 Camerino, Italy}
\date{Received \today}
\maketitle
\widetext
\begin{abstract}
A new master equation performing isotropic phase-number squeezing is suggested.
The phase properties of coherent superpositions are analyzed when the state
evolves in presence of a bath with fluctuations squeezed in this isotropic
way. We find that such a reservoir greatly improves persistence of coherence
with respect to either a customary thermal bath, or to an anisotropically
squeezed phase-sensitive bath.
\end{abstract}
\narrowtext

\section{Introduction}
\label{intro}

Recently, much attention has been focused on quantum interference
effects for superpositions of macroscopically distinguishable
states~\cite{kn:wamil,kn:yusto,kn:kenw,kn:mimto,kn:brisu,kn:buz},
with attempts of observing nonclassical
features at a macroscopic level~\cite{kn:leg,kn:legg,kn:wight}.
Considerable effort has been devoted to the investigation of
the influence on macroscopic superpositions due to dissipation, which rapidly
destroys quantum coherence, and makes quantum effects non detectable in
practice~\cite{kn:miw,kn:caleg,kn:sawa,kn:miho,kn:miwa,kn:dami}.

With the aim of reducing the effect of dissipation on a
macroscopic superposition, Kennedy and Walls~\cite{kn:kenw} have studied the
time evolution of an initial superposition of coherent states for a single
mode of the field interacting with a bath having squeezed
fluctuations~\cite{kn:nota}. They have shown that when the
fluctuations are squeezed in the right quadrature a squeezed bath becomes
more efficient than a thermal one in preserving interference
fringes in the output photocurrent of a homodyne detector.
More recently, Bu\v{z}ek, Kim and Gantsog~\cite{kn:buz} have investigated
the phase properties of quantum superpositions of two coherent states
under squeezed amplification, showing that a suitable phase-sensitive
amplifier is able to preserve the phase
distribution of the input state (Schr\"odinger-cat state~\cite{kn:scro}).

In Refs.~\cite{kn:kenw} and \cite{kn:buz} a superposition of two coherent
components is considered as a test quantum superposition. In this case the
quantum state itself determines a privileged direction---the line joining the
two peaks in the complex plane---which selects the quadrature suited to
squeezing. Nevertheless, superpositions of macroscopically distinguishable
states can be formed by more than two states. For example, \hfill
during the time evolution of
\newline \noindent \vskip 3.3truecm \noindent
an  initial coherent state in a Kerr medium~\cite{kn:yusto}, one can
have superpositions of three, four, and even more coherent states.
For this
reason, in this paper we suggest a new master equation for ``isotropic''
phase-number squeezing, which is much more efficient in preserving
coherence of any general superposition compared to an anisotropic squeezed
bath. We analyze coherence by observing the phase (quasi)probability
that is marginal of the Wigner function~\cite{kn:gakni,kn:tan,kn:bukig}
\begin{equation}
P(\phi)=\int_0^{\infty} r W(r e^{i\phi},r e^{-i\phi}) dr\;,
\label{eq:tpd}
\end{equation}
where the Wigner function is defined by
\begin{eqnarray}
W(\alpha,\alpha^{\ast})=\int{{d^2\lambda}\over{\pi^2}}e^{-\lambda\alpha
^{\ast}+{\lambda}^{\ast}\alpha }\,\hbox{Tr}\left\{\hat\rho e^{\lambda
a^{\dagger}-{\lambda^{\ast}}a} \right\} \;,
\label{eq:wigne}
\end{eqnarray}
and $\hat{\rho}$ is the density matrix of the system.
We will compare numerical results for the time evolution of an
initial superposition state in the cases of isotropic, directional,
and vanishing squeezing, concluding that the isotropic squeezing is
much more effective in preserving the peak structure of the phase
distribution than the other two cases.

This paper is organized as follows: In Section~\ref{master} we introduce the
model and the master equation. In Sec.~\ref{persi} we show the results of
numerical integration in terms of $P(\phi)$ and of the Wigner function.
Section~\ref{conclu} concludes the article with a short
discussion and some remarks.

\section{The master equation}
\label{master}

The master equation for the reduced density operator of a single field mode
in a squeezed bath can be derived from the knowledge of the correlation
functions of the bath operators~\cite{kn:gardi}. In the interaction picture
and in the rotating wave approximation one obtains \cite{kn:kenw}
\begin{eqnarray}
{d\hat{\rho}\over dt} & = &
\gamma(N+1)(2a\hat{\rho} a^{\dagger}-
a^{\dagger} a\hat{\rho}-\hat{\rho} a^{\dagger} a) \nonumber\\
& & \mbox{} + \gamma N(2a^{\dagger}\hat{\rho} a-
aa^{\dagger}\hat{\rho}-\hat{\rho} aa^{\dagger})
\nonumber \\
 & & \mbox{} -
\gamma M(2a^{\dagger}\hat{\rho} a^{\dagger}-a^{\dagger}
a^{\dagger}\hat{\rho}-\hat{\rho} a^{\dagger} a^{\dagger}) \nonumber\\
& & \mbox{} -\gamma M^{\ast}(2a\hat{\rho} a-aa\hat{\rho}
-\hat{\rho} aa)\;,
\label{eq:mequni}
\end{eqnarray}
where $a$,$a^{\dagger}$ are the boson annihilation and creation operators of
the mode, $\gamma$ is the damping constant, and $M=|M|e^{i\psi}$ is the
squeezing complex parameter satisfying $|M|^2\leq N(N+1)$. For $M=0$ the
reservoir reduces to an usual thermal bath and $N$ becomes the mean number of
thermal photons; for $|M|^2 =N(N+1)$ the squeezing is maximum. The master
equation (\ref{eq:mequni}) describes a situation in which the noise transferred
to the system increases quadrature fluctuations in one direction
more rapidly than in the other ones. The choice of the squeezing direction is
determined {\it a priori} by the phase $\psi$ of the squeezing
parameter $M$. In this sense the master equation~(\ref{eq:mequni})
represents an ``unidirectionally'' squeezed bath.

In order to obtain an ``isotropic'' squeezing we modify Eq.~(\ref{eq:mequni})
in such a way that the phase of $M$ is dynamically shifted as a function on
the phase of the field. For highly excited states $\hat\rho$ the phase
factor of the state can be approximately given by the following
expectation value
\begin{equation}
e^{\pm i\phi}\simeq\hbox{Tr}[\hat e^{\mp}\hat\rho ]\;,\label{exp}
\end{equation}
where $\hat e^-$ and $\hat e^+\equiv (\hat e^-)^{\dag}$ are the shift
operators
\begin{equation}
\hat{e}^- = (aa^{\dagger})^{-1/2}a\;,\qquad
\hat{e}^{+} = a^{\dagger}(aa^{\dagger})^{-1/2}\;,
\label{eq:epm}
\end{equation}
acting on the Fock space as follows
\begin{equation}
\hat{e}^{+}|n\rangle = |n+1\rangle\;,\qquad
\hat{e}^{-}|n\rangle = |n-1\rangle\;.
\label{eq:Fock}
\end{equation}
This suggests using $\hat e^{\mp}$ as dynamical phase factors in the master
equation. As the phase of squeezing rotates at double frequency than the
average field, the shift operators should appear in pairs in
the squeezing part of the Liouvillian (\ref{eq:mequni}). This last observation,
along with the requirement of an isometrical master equation (which must
preserve normalization of $\hat\rho$) leads us to consider the following
substitutions in the squeezing term of Eq. (\ref{eq:mequni})
\begin{equation}
a^{\dagger} \longrightarrow \hat{e}^{-}a^{\dagger}\;,\qquad
a \longrightarrow a\hat{e}^{+}\;.
\label{eq:sub}
\end{equation}
Taking into account the following operator identities
\begin{equation}
\hat e^- a^{\dagger} =a\hat e^+ =\sqrt{\hat n+1}\;,
\label{eq:iden}
\end{equation}
the substitutions (\ref{eq:sub}) suggest changing the master equation
(\ref{eq:mequni}) into the form
\begin{eqnarray}
{d\hat{\rho}\over dt} & = &
\gamma(N+1)(2a\hat{\rho} a^{\dagger}-
a^{\dagger} a\hat{\rho}-\hat{\rho} a^{\dagger} a) \nonumber\\
& & \mbox{} + \gamma N(2a^{\dagger}\hat{\rho} a-
aa^{\dagger}\hat{\rho}-\hat{\rho} aa^{\dagger})
\nonumber \\
& & \mbox{} - 2\gamma \Re(M)[2\sqrt{\hat n+1}\hat{\rho}\sqrt{\hat n+1}
\nonumber \\
 & & \mbox{} \; \; \; \; \; \; \; \; \; \; \; \; \; \; \; \; \; \; \;
-(\hat n+1)\hat{\rho}-\hat{\rho}(\hat n+1)]\;.
\label{eq:meq}
\end{eqnarray}
The master equation (\ref{eq:meq}) here derived in a heuristic way could
represent a feedback-driven thermal bath which detects the phase of the
state dynamically. As we will see, the relaxation (\ref{eq:meq}) is very
effective in preserving coherence, and thus a more fundamental
derivation of (\ref{eq:meq}) is motivated (work is in progress along this
line). Notice that the stationary solution of Eq. (\ref{eq:meq}) is
still the thermal distribution, but the decay of the off-diagonal terms
is achieved for longer times.

The time evolution of an initial coherent state
$|\alpha_0\rangle=|4.0\rangle$ is given in Fig.~\ref{f:1} for two
different values of
$M$. Here the Wigner function and the phase distribution $P(\phi)$
are plotted at $t=0.1\,\gamma^{-1}$ for $N=10$ and $M=\pm\sqrt{N(N+1)}$
[actually, $M$ now becomes a real parameter, because its imaginary part
does not
contribute in (\ref{eq:meq})]. It is evident that the effect of the isotropic
squeezing corresponds to phase-squeezing for $M>0$, and to number-squeezing
for $M<0$. Due to dissipation, in addition to squeezing the Wigner function
also moves slightly towards the origin of the phase space. In Sect.~\ref{persi}
we will study the time evolution of a superposition of three coherent states
(generated via Kerr effect), and we will analyze the persistence of
phase coherence of the state.

\section{Persistence of phase coherence}
\label{persi}

The Kerr nonlinearity in quantum optics is probably the best candidate
to produce quantum superpositions of macroscopically distinguishable
states. In Ref.~\cite{kn:yusto}, Yurke and Stoler investigated the time
evolution of an initial coherent state propagating through such an
\lq\lq amplitude-dispersive\rq\rq\ medium which is described by the
Hamiltonian
\begin{equation}
H=\omega(a^{\dagger}a)+\Omega(a^{\dagger}a)^2\;.
\label{eq:hamyu}
\end{equation}
They showed that the state vector is periodic with period $2\pi/\Omega$
and that during a period the state evolves passing through symmetrical
superpositions of $k$ coherent states at times $t=\pi/k\Omega$ (for not
too large $k$).

For a superposition of two coherent states (symmetrical
with respect to the origin) there is no difference between isotropic
and anisotropic squeezing. Therefore we are interested in the
simplest superposition with more than two
components, namely the state obtained for $t=\pi/3\Omega$
\begin{eqnarray}
|\phi\rangle_3 & = & \frac{1}{\sqrt{3}}
\left[e^{-i\pi/6}|\alpha_0e^{i\pi/3}\rangle+e^{i\pi/2}|-\alpha_0\rangle
\right.
\nonumber \\
 & & \mbox{} \; \; \; \; \; \; \; +\left.e^{-i\pi/6}|\alpha_0e^{-i\pi/3}\rangle
\right]\;.
\label{eq:suptre}
\end{eqnarray}

In Fig.~\ref{f:2} the contour plot of the Wigner function corresponding
to the state (\ref{eq:suptre}) is given for $\alpha_0=-4.0$. In the
following the phase properties of the time evolution of the state
$|\phi\rangle_3$ according to the master equation (\ref{eq:meq}) are
investigated, and the results are compared with those obtained with the same
initial state for thermal bath---Eq.~(\ref{eq:mequni}) with $M=0$---and
for anisotropically squeezed bath---Eq.~(\ref{eq:mequni}) with $M^2=N(N+1)$.
The time evolution has been integrated numerically, and the Wigner function
has been obtained using fast Fourier transform techniques.

In Fig.~\ref{f:2} the time evolved Wigner function and $P(\phi)$ are given
for isotropic squeezing $M=\sqrt{N(N+1)}$ and $N=30$; in Fig.~\ref{f:3} and
Fig.~\ref{f:4} for comparison the two distributions are plotted at the
same evolved times, but for anisotropic squeezing and thermal bath,
respectively. The effectiveness of Eq. (\ref{eq:meq}) in preserving
coherence is evident. For isotropic squeezing the phase distribution
$P(\phi)$ survives for much longer times, and the three peaks remain
identical each other for all times. On the contrary, for anisotropic squeezing
the central peak is the only one surviving at long times, whereas the other
two ones are smeared out. Notice that the height of the central peak for
anisotropic squeezing is greater than that of the three peaks for
isotropic squeezing at the same times: such survival of
coherence in the directional case, however, is fortuitous, being strongly
dependent on the direction of squeezing with respect to the angular shape
of the Wigner function in the complex plane.

In concluding this section, we notice that the effect of vacuum
component of the squeezed bath in washing out quantum interference is
stronger for lower values of $N$ and $M=\sqrt{N(N+1)}$. In Fig.
\ref{f:5} this is shown for $N=3$, where at $\gamma t=0.01$ the
interference patterns are weakly visible, whereas at $\gamma t=0.02$
they are already totally absent. These plots should be compared with
those in Fig. \ref{f:2}, which correspond to much longer times.

\section{Summary and conclusions}
\label{conclu}

In this paper we have suggested a new master equation which squeezes
states isotropically in the complex plane. This novel kind of squeezing
turns out to be very effective in increasing the persistence time of
coherence, independently on the quantum state. This suggests an improved
scheme of detection and generation of Schr\"odinger-cat states, based
on isotropically squeezed reservoirs. The master equation of the
isotropically squeezed bath has been derived heuristically: a suitable
feedback mechanism should now be envisaged, which supports this new type of
dissipative dynamics.

\begin{figure}
\caption{The effect of isotropic squeezing
[master equation~(\protect\ref{eq:meq})] for positive and negative
values of $M$ on a coherent state with $\alpha _0=4$. Contour plots of the
Wigner function and $P(\phi)$ distribution evolved in time for $N=10$ and
$|M|=\protect\sqrt{N(N+1)}$.}
\label{f:1}
\end{figure}
\begin{figure}
\caption{Contour plots of the Wigner function and $P(\phi)$ distribution
evolved in time for initial state~(\protect\ref{eq:suptre}), and for isotropic
squeezing [master equation~(\protect\ref{eq:meq})]. Here $N=30$,
$M=\protect\sqrt{N(N+1)}$ and $\alpha _0=-4$.}
\label{f:2}
\end{figure}
\begin{figure}
\caption{As in Fig.~\protect\ref{f:2}, but for anisotropic squeezing
[master equation~(\protect\ref{eq:mequni})].}
\label{f:3}
\end{figure}
\begin{figure}
\caption{As in Fig.~\protect\ref{f:2}, but for customary thermal bath
($M=0$).}
\label{f:4}
\end{figure}
\begin{figure}
\caption{As in Fig.~\protect\ref{f:2}, but for $N=3$ and different
times.}
\label{f:5}
\end{figure}
\end{document}